\documentclass{article} 
\usepackage[final]{colm2025_xllmworkshop}

\usepackage{microtype}
\usepackage{hyperref}
\usepackage{url}
\usepackage{booktabs}
\usepackage{graphicx}
\usepackage{tabularx}
\usepackage{enumitem}
\usepackage{float}

\usepackage{lineno}

\definecolor{darkblue}{rgb}{0, 0, 0.5}
\hypersetup{colorlinks=true, citecolor=darkblue, linkcolor=darkblue, urlcolor=darkblue}

\newcommand{\sysname}{\textsc{CopilotLens}}

\title{Beyond Autocomplete: Designing \sysname{} Towards Transparent and Explainable AI Coding Agents}

\author{%
  Runlong Ye\\
  Computer Science, University of Toronto\\
  Toronto, Ontario, Canada\\
  \texttt{harryye@cs.toronto.edu}
  \And
  Zeling~Zhang\\
  Computer Science, University of Toronto\\
  Toronto, Ontario, Canada\\
  \texttt{zeling.zhang@mail.utoronto.ca}\\
  \And
  Boushra~Almazroua\thanks{Work done during research internship at the University of Toronto.}\\
  King Abdullah University of Science and Technology\\
  Thuwal, Jeddah, Saudi Arabia\\
  \texttt{boushra.almazroua@mail.utoronto.ca}\\
  \AND
  Michael Liut\\
  Mathematical and Computational Sciences, University of Toronto Mississauga\\
  Mississauga, Ontario, Canada\\
  \texttt{michael.liut@utoronto.ca}
}

\begin{document}

\ifcolmsubmission
\linenumbers
\fi

\maketitle

\begin{abstract}
AI-powered code assistants are widely used to generate code completions, significantly boosting developer productivity. However, these tools typically present suggestions without explaining their rationale, leaving their decision-making process inscrutable. This opacity hinders developers’ ability to critically evaluate outputs, form accurate mental models, and calibrate trust in the system. To address this, we introduce \sysname{}, a novel interactive framework that reframes code completion from a simple suggestion into a transparent, explainable interaction. \sysname{} operates as an explanation layer that reconstructs the AI agent’s ``thought process'' through a dynamic, two-level interface. The tool aims to surface both high-level code changes and the specific codebase context influences. This paper presents the design and rationale of \sysname{}, offering a concrete framework and articulating expectations on deepening comprehension and calibrated trust, which we plan to evaluate in subsequent work.
\end{abstract}

\section{Introduction}


AI-powered assistants like GitHub Copilot\footnote{\href{https://github.com/features/copilot}{https://github.com/features/copilot}} and more intelligent coding agents like Cursor\footnote{\href{https://www.cursor.com/}{https://www.cursor.com/}} and WindSurf\footnote{\href{https://windsurf.com/}{https://windsurf.com/}} are now essential tools in modern software development, evolving from simple code completion to autonomously executing complex, project-wide tasks \cite{liang2024, weisz2025}. This trend is accelerating with the rise of asynchronous agents like Google's Jules\footnote{\href{https://jules.google/}{https://jules.google/}}, the rebranded OpenAI Codex\footnote{\href{https://openai.com/index/introducing-codex/}{\href{https://openai.com/index/introducing-codex/}{https://openai.com/index/introducing-codex/}}}, and Cursor's web-based agents\footnote{\href{https://www.cursor.com/blog/agent-web/}{https://www.cursor.com/blog/agent-web/}}, which abstract the programmer further away from the code. However, this utility comes at the cost of clarity \cite{vasconcelos2025, brachman2025personalized}. Mainstream assistants prioritize suggestion speed over reasoning, leaving their decision-making process inscrutable and forcing programmers to reverse-engineer the AI's intent. This opacity is a significant usability challenge that hinders the development of accurate mental models, suppresses critical evaluation of generated code \cite{kazemitabaar2025engagement, mozannar2024}, and fosters a fragile, uncalibrated trust \cite{vaithilingam2022}.

To address this critical gap between suggestion and comprehension, we argue for a paradigm shift toward explainable autocomplete \cite{yan2024ivie, brachman2025personalized} and introduce \sysname{}, an interactive explanation layer designed to reconstruct the ``thought process'' of an AI coding agent retrospectively. It reframes code completion from an isolated output into a transparent, reviewable event by surfacing the agent's plan, contextual evidence, and followed conventions. 

This brief investment in transparency is designed to serve distinct developer needs. For \emph{novice} student programmers, it transforms AI-assisted coding into a tangible learning experience, clearly exposing the underlying reasoning process to foster accurate mental models \cite{james2023weird}. For \emph{professional} developers, it aims to enable the rapid verification of complex multi-file suggestions, helping to prevent the accidental integration of subtle yet costly errors that prior large-scale studies have shown to be a persistent challenge \cite{liang2024}. More concrete use cases can be found in Appendix \ref{sec:use-case}.

The central design challenge we address is: \emph{\textbf{How can the obscure reasoning process of an AI coding agent be redesigned into a transparent process that bridges the gap between an agent's internal logic and a developer's mental model, thereby supporting critical evaluation and fostering calibrated trust?}}

We present \sysname{} as an initial design response to this question. We make the following contributions: (1) we articulate the challenges arising from the lack of transparency in current code assistants and formulate a set of design goals for creating more comprehensible tools; and (2) we introduce a dynamic, two-level explanation framework that exposes an agent's reasoning process, from high-level file modifications to the specific codebase influences behind a single change. By illuminating the agent's decision-making process, this work outlines a design framework and a research agenda for evaluating next-generation code assistants that not only write code, but also foster deeper understanding and more calibrated trust.

\section{Related Work}
Developers have long grappled with the cognitive challenges of understanding large and complex software systems. To manage this, they have historically relied on a variety of strategies to build and maintain a mental model of the code. Foundational techniques include comprehensive software documentation \cite{402076, 10.1145/1134285.1134355}, visual notations like UML to make system architecture tangible \cite{proceedings2021074013}, abstraction to hide implementation details \cite{385970}, and program slicing to isolate relevant code \cite{10.5555/869354}. While effective, these traditional methods were designed to help humans comprehend \emph{existing}, \emph{human-written} codebases. The rise of AI-powered assistants has fundamentally altered this landscape, introducing a new kind of comprehension gap that these static methods are ill-equipped to address.

LLM-powered code assistants have rapidly transformed productivity gains for developers at every skill level \cite{weber2024, mozannar2024}. Yet this productivity surge comes with a significant trade-off: opacity. Current LLM code assistants contribute code solutions without revealing the reasoning behind them \cite{husein2024, ferdowsi2024}. This creates a new, active comprehension challenge: one not about understanding a static artifact, but about understanding the inscrutable \emph{reasoning process} of an AI agent as it generates ever more complex programs. As a result, users often engage with these systems as ``black boxes'' and settle for a ``good-enough'' understanding \cite{oldenburg2025}. This dynamic obstructs the formation of accurate mental models, discourages critical evaluation, and promotes uncalibrated trust \cite{lee2025criticalthinking, wang2024}.  

Opacity introduces not only usability friction and trust concerns but also deeper cognitive burdens. Developers must often reverse-engineer the AI's intent post hoc, expending cognitive effort to infer why certain completions were offered \cite{Weisz2021, weisz2025, brown2024}. This cognitive overhead is particularly challenging for novices, who may lack sufficient domain expertise to recognize subtle errors or inappropriate suggestions \cite{kumar2024,viktoria2025, Kazemitabaar2024}. Over-reliance becomes common—and sometimes even conscious, among users \cite{james2023weird}, as they can either blindly accept AI output or unknowingly propagate flawed code.

In response to these challenges, explainable AI (XAI) and Human-centered XAI (HXAI) have sought to enhance transparency and redefine explainability for LLMs \cite{sun2022genai, kim2025}. Several promising paradigms have emerged. Token-level attribution allows users to trace model outputs to influential training examples, aiding provenance reasoning \cite{li2024attributor}. Uncertainty visualization approaches offer complementary insights: by surfacing model confidence or expected edit likelihoods, they help users triage which suggestions warrant additional scrutiny \cite{bhatt2022uncertainty, vasconcelos2025}.  However, these signals remain orthogonal to explaining the model's internal reasoning process, leaving the ``why'' behind generation opaque. \cite{Vasconcelos2023} emphasizes that different forms of explanation vary in their effectiveness, with explanations that make the AI’s mistakes more salient being particularly effective at reducing overreliance. For example, self-consistency checks, where models verify their own outputs across multiple reasoning passes, have shown promise for detecting hallucinations or factual inconsistencies \cite{leiser2024, cheng2024}. 

Despite such advances, current methods remain largely output-centric. They visualize or score the final artifact without exposing the reasoning that produced it, leaving users to reconstruct the logic from surface-level signals alone \cite{Miller2019}. Recent exploratory work has begun to suggest pathways for surfacing intermediate planning structures to bridge this divide \cite{yen2023coladder}. \sysname{}, on the other hand, seeks to further close this gap by reframing code generation as an inspectable event. Rather than merely evaluating what the model produced, we seek to make visible how it arrived at that output, providing developers with actionable insight into the model's decision-making process and enabling richer, more trustworthy collaboration.

\section{Design Goals}
\label{subsection:design_goals}

\begin{figure}[ht]
    \centering
    \includegraphics[width=\linewidth]{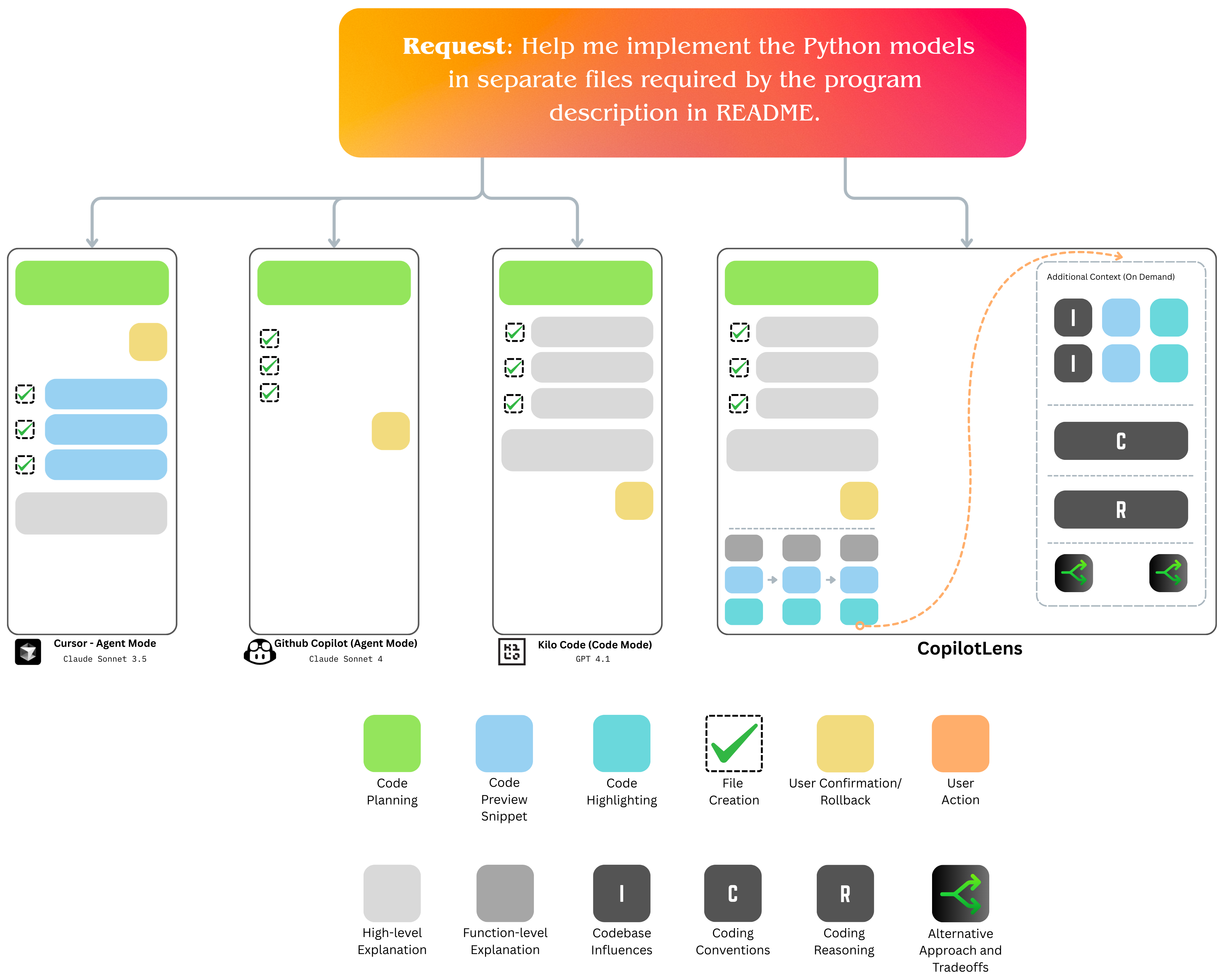}
    \caption{A Comparison of AI Coding Agent Interaction Panels. Mainstream tools like Cursor, GitHub Copilot, and open-source coding agent Kilo Code (left) present a final code artifact with minimal process visibility and explanations. In contrast, \sysname{} (right) reconstruct the agent's ``thought process'' through a two-level explanation, surfacing important codebase influence (I), coding convention (C), coding reasoning (R), and alternative approach to consider, to support critical evaluation and foster calibrated trust.}
    \label{fig:coding-agent-comparison}
\end{figure}

AI-powered coding assistants that prioritize speed of suggestion over clarity of reasoning create significant usability challenges that hinder effective human-AI collaboration \cite{liang2024, mozannar2024, vasconcelos2025, Terragni2025}. Mainstream AI tools, such as GitHub Copilot and Cursor, typically present a final code artifact with minimal insight into the agent's decision-making process (Figure \ref{fig:coding-agent-comparison}, left). This opacity forces programmers to reverse-engineer the AI's intent \cite{sergeyuk2025human}, which can lead to misaligned expectations, suppressed critical evaluation of generated code, and fragile, uncalibrated user trust \cite{vaithilingam2022}. Drawing from these established challenges, we formulate three design goals to guide the development of a more transparent and comprehensible code generation experience.

\subsection{Challenge: Bridging the Gap Between AI Opacity and Developer Mental Models}

A primary challenge with AI coding assistants is that developers lack accurate mental models of how the agent reaches its solutions \cite{vaithilingam2022, barke2023grounded}. Suggestions often feel ``out-of-the-blue'' because the agent's high-level plan, interpretation of user requests, and considered context remain hidden \cite{ross2023}. Consequently, developers interact with these tools as if they are inscrutable, overly confident partners.

The opacity of AI coding assistants actively reinforces flawed mental models and promotes ineffective developer behavior. This problem is worsened by minimalist interfaces that, while boosting productivity, hide the AI’s reasoning by concealing which files it consulted or the steps it took. Forced to guess at the agent's capabilities, programmers are frequently inaccurate \cite{vaithilingam2022}. These misunderstandings can lead to poor prompts or the blind acceptance of flawed code, stemming from an overestimation of the AI's contextual understanding and resulting in a fragile, uncalibrated trust \cite{murillo2024understanding}.

\emph{\textbf{Design Goal G1}: Reveal the Agent's Thought Process.} A system should not just provide a final solution; instead, it should surface the breakdown of actions the agent took to arrive at that solution, making its process explicit and reviewable to bridge the user's mental model gap.

\subsection{Challenge: Providing Deep Context to Scaffold Critical Evaluation}

Simply revealing an agent's action sequence is insufficient for promoting deep comprehension \cite{wood1976role}. Effective scaffolding must connect AI-generated code changes explicitly to the broader codebase context. Recent AI-driven approaches, such as the ephemeral UIs in computational notebooks, dynamically provide in-context explanations that enhance code comprehension \cite{biscuit2024scaffolding}. Research shows that integrating architectural, API, and file-level context is critical for improving a user's ability to evaluate generated suggestions \cite{wang2025towards}.

Furthermore, effective scaffolds should provide layered, contextual rationales rather than isolated hints \cite{hou2024parsons}, and must adaptively fade as developers gain expertise to avoid diminishing user autonomy \cite{vangog2020assistance}. Hierarchical frameworks like CoLadder demonstrate that revealing intermediate planning steps improves comprehension by aligning the generated code with user intentions \cite{yen2023coladder}. Similarly, tools like CodeCompass show that automatically surfacing relevant snippets from a repository aids developers in evaluating unfamiliar code \cite{agrawal2024codecompassstudychallenges}.

\emph{\textbf{Design Goal G2}: Support Informed Critical Evaluation.} The system should scaffold developers' critical evaluation by providing adaptive, deep contextual explanations linking specific code changes to project architecture, conventions, and design trade-offs.

\subsection{Challenge: Making AI Reasoning Transparent to Foster Calibrated Trust}
To trust AI-assisted development, developers need clear evidence that the system is not only correct but also makes logical decisions based on the project's context. Empirical findings by \cite{brown2024} demonstrate that developers’ trust in AI code suggestions significantly improves when they understand the assumptions and contextual reasoning behind the generated outputs, rather than relying solely on raw accuracy. Evaluations of contemporary AI-based assistants further show that generated methods frequently overlook project-specific constraints, highlighting the limitations of post-hoc verification as a sole mechanism for fostering trust \cite{corso2024}. 

To address this gap, recent research has proposed ``trust affordances'', such as suggestion quality indicators and explicit usage analytics, enabling developers to form more accurate mental models of AI behavior by transparently communicating how suggestions align with established project norms and context \cite{wang2024}.

\emph{\textbf{Design Goal G3}: Make AI Reasoning Tangible and Verifiable.} The interface should explicitly present tangible evidence of the AI’s reasoning, clearly linking each suggestion to the specific files, coding conventions, and architectural patterns used in its generation, thereby enabling developers to assess alignment with project standards immediately.

\section{\sysname{} Probe: An Interactive and Modular Explanation Layer}
\begin{table}[ht]
\raggedright
\small
\begin{tabularx}{\textwidth}{X X X X}
\toprule
\textbf{Identified Challenge in AI Code Generation} 
  & \textbf{Manifestation in Current Tools (e.g., Copilot, Cursor)} 
  & \textbf{\sysname{} Design Goal} 
  & \textbf{Corresponding Feature(s)} \\
\midrule

\emph{Challenge 1: Opaque Reasoning Process \& Misaligned Mental Models} 
  & Suggestions appear “out-of-the-blue” with no insight into the model’s plan or decision points, leading to confusion \cite{vaithilingam2022}. 
  & G1: Expose the Agent’s Thought Process. 
  & Level 1 Explanations: Streaming, sequential display of file modifications and interactive code highlighting. \\[1ex]
\\
\emph{Challenge 2: Lack of Contextual Understanding \& Suppressed Critical Inquiry} 
  & Speed and fluency of suggestions discourage critical evaluation for subtle flaws or alternatives \cite{mozannar2024}. 
  & G2: Support Informed Critical Evaluation. 
  & Level 2 Explanations (On-Demand): Analysis of codebase influences, coding conventions, and alternative implementations. \\[1ex]
\\
\emph{Challenge 3: Making AI Reasoning Transparent to Foster Calibrated Trust} 
  & Users cannot directly assess suggestion reliability, as correctness is hidden, making trust fragile and uncalibrated \cite{brown2024, wang2024}. 
  & G3: Make Model Confidence and Correctness Tangible. 
  & Level 2 Explanations (On-Demand): Explicitly linking code changes to project-specific artifacts (e.g., other files, documentation) to demonstrate process integrity. \\

\bottomrule
\end{tabularx}
\caption{Design challenges, their manifestations in current AI code assistants, \sysname{}’s three design goals, and the features that realize each goal.}
\label{tab:design-challenges}
\end{table}

To address our design goals, we designed and implemented \sysname{} (Figure \ref{fig:coding-agent-comparison}, right). \sysname{} is an interactive explanation layer built on top of an existing open-source coding agent, Kilo Code\footnote{https://kilocode.ai/}. At its core, \sysname{} intercepts the output of a coding agent \emph{after} a task is complete to analyze and explain its actions. This design is guided by a specific rationale and is structured into a dynamic, two-level framework, which we detail below.

\subsection{Design Rationale: Post-Hoc, Model-Agnostic Explanation}

The decision to perform post-hoc analysis is a pragmatic one. While direct, on-the-fly model interpretability is a primary goal of the broader XAI community, current techniques are not yet practical for our target workflow \cite{zhao2023explainabilitylargelanguagemodels}. State-of-the-art methods often require privileged ``white-box'' model access, which presents a difficult trade-off: it pushes developers to interact with open-weight models (i.e., LLama, Qwen, Mistral), while valuable for research, often underperform against leading proprietary APIs (i.e., GPT-5, Gemini 2.5,  Claude 4) in complex coding tasks \cite{jimenez2024swebenchlanguagemodelsresolve}. This would force an undesirable choice between developer productivity and model explainability. Beyond this, such techniques risk introducing significant latency or producing low-level, token-centric explanations that are difficult to map to a developer's high-level goals \cite{burns2024discoveringlatentknowledgelanguage}. Such low-level signals often fail to provide meaningful insight into an agent's broader strategy, a long-standing challenge in XAI \cite{Miller2019, jain2019attentionexplanation}. 

By instead reconstructing the agent’s ``thought process'' from its final outputs, \sysname{} remains model-agnostic, allowing developers to use the most powerful coding agents available while still benefiting from a transparent, actionable explanation layer.

\subsection{A Dynamic, Two-Level Explanation Framework} 
\vspace{-.5em}

\sysname{}'s central design principle is a two-level dynamic explanation framework designed to manage the inherent trade-off between informational quality and cognitive load, a key challenge in explainable AI (XAI) \cite{Miller2019,doshivelez2017rigorousscienceinterpretablemachine,sweller1988cognitive}. It is designed as a research probe to investigate how programmers interact with AI-generated code when the ``black box'' is opened \cite{sun2022genai}. This two-level design is structured using the concepts of \emph{explanatory} versus \emph{exploratory} user interfaces (XUIs) \cite{chromik2021human}.
 \begin{itemize}
     \item \textbf{Level 1} functions as an \emph{explanatory XUI}. It is designed for immediate, ``at-a-glance'' awareness, presenting a concise, post-hoc summary of the agent's actions. Its purpose is to quickly answer the question: ``What just happened?''.

    \item \textbf{Level 2} functions as an \emph{exploratory XUI}. It is an on-demand, user-driven environment for deep investigation. Its purpose is to allow a developer to probe the rationale behind a specific change, answering the question: ``Why was it done this way?''.
 \end{itemize}
 
By separating these two modes of interaction, the system supports both rapid sensemaking and deep reflective analysis, allowing developers to engage with the explanation at the level of detail appropriate for their current task.

\begin{figure}[ht]
    \centering
    \includegraphics[width=\linewidth]{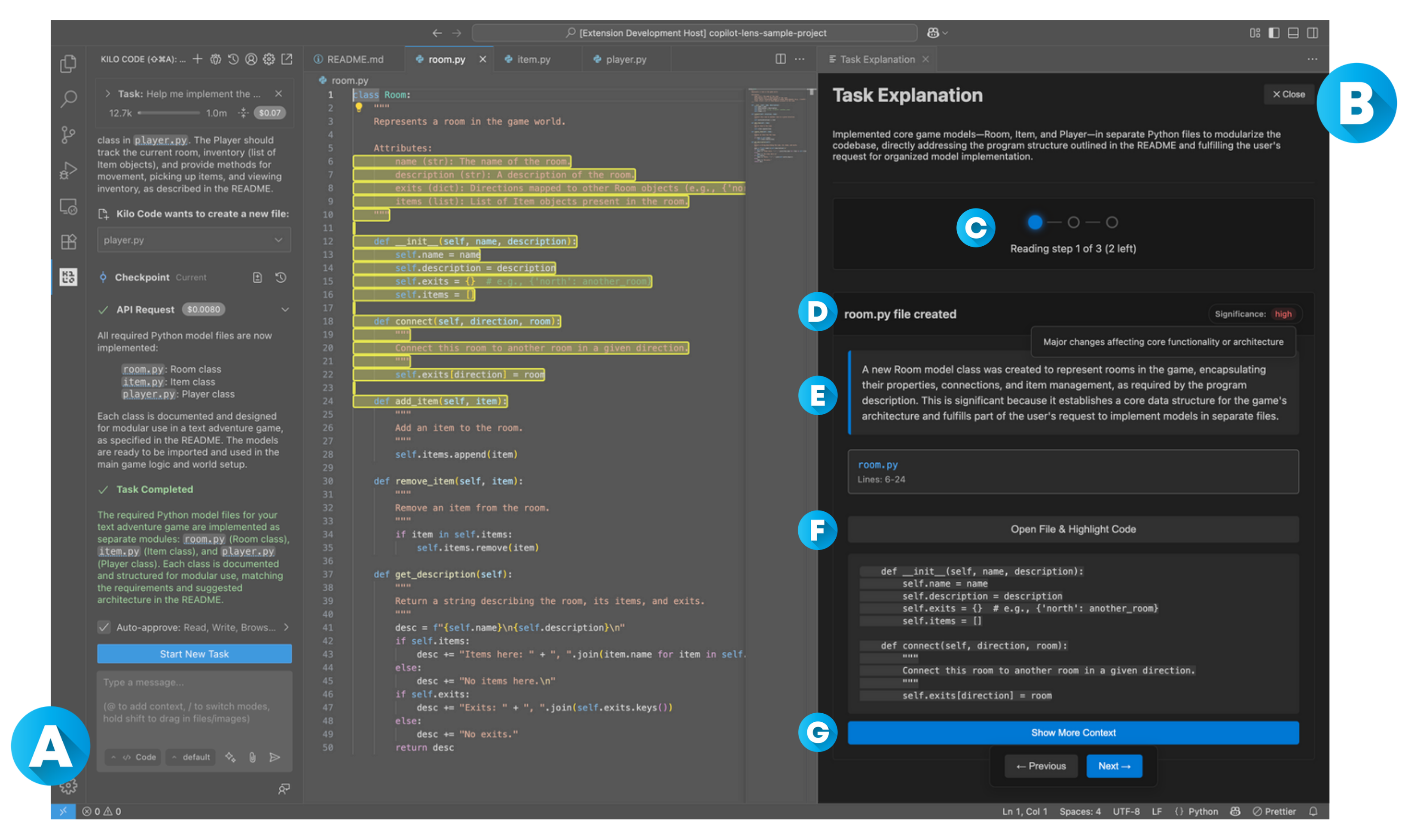}
    \caption{Main Interface for \sysname{}. (A) \sysname{} interface is presented after the AI coding agent has completed its assigned task. (B) The main explanation view for \sysname{}, which provides a post-hoc, two-level analysis of the agent’s actions. The default Level 1 explanation offers a summary of what was changed, including (C) a visual navigator to step through each file modification, (D) the title of the current modification and its significance, (E) an AI-generated summary of the change's purpose and significance, and (F) a preview of the implemented code and interactive highlights. (G) The user can click to trigger the on-demand Level 2 analysis, which provides an extended explanation.}
    \label{fig:copilot-lens-main-interface}
\end{figure}

\subsection{Level 1: Post-Hoc Explanatory Summaries of Modifications (\autoref{fig:copilot-lens-main-interface})}
\vspace{-.5em}
In response to \emph{G1 (Reveal the Agent's Process)}, \sysname{} presents a post-hoc explanatory summary of the agent's modifications.  This initial level of explanation is designed for immediate situational awareness, automatically providing a per-file overview of the changes made. It features a side panel that sequentially displays each modified file with a concise summary of its purpose and significance, enabling rapid comprehension. Clicking on a summary navigates the user to the corresponding code changes (i.e., modified functions) in the code editor, creating a direct visual link between the explanation and the implementation.

\begin{figure}[ht]
    \centering
    \includegraphics[width=\linewidth]{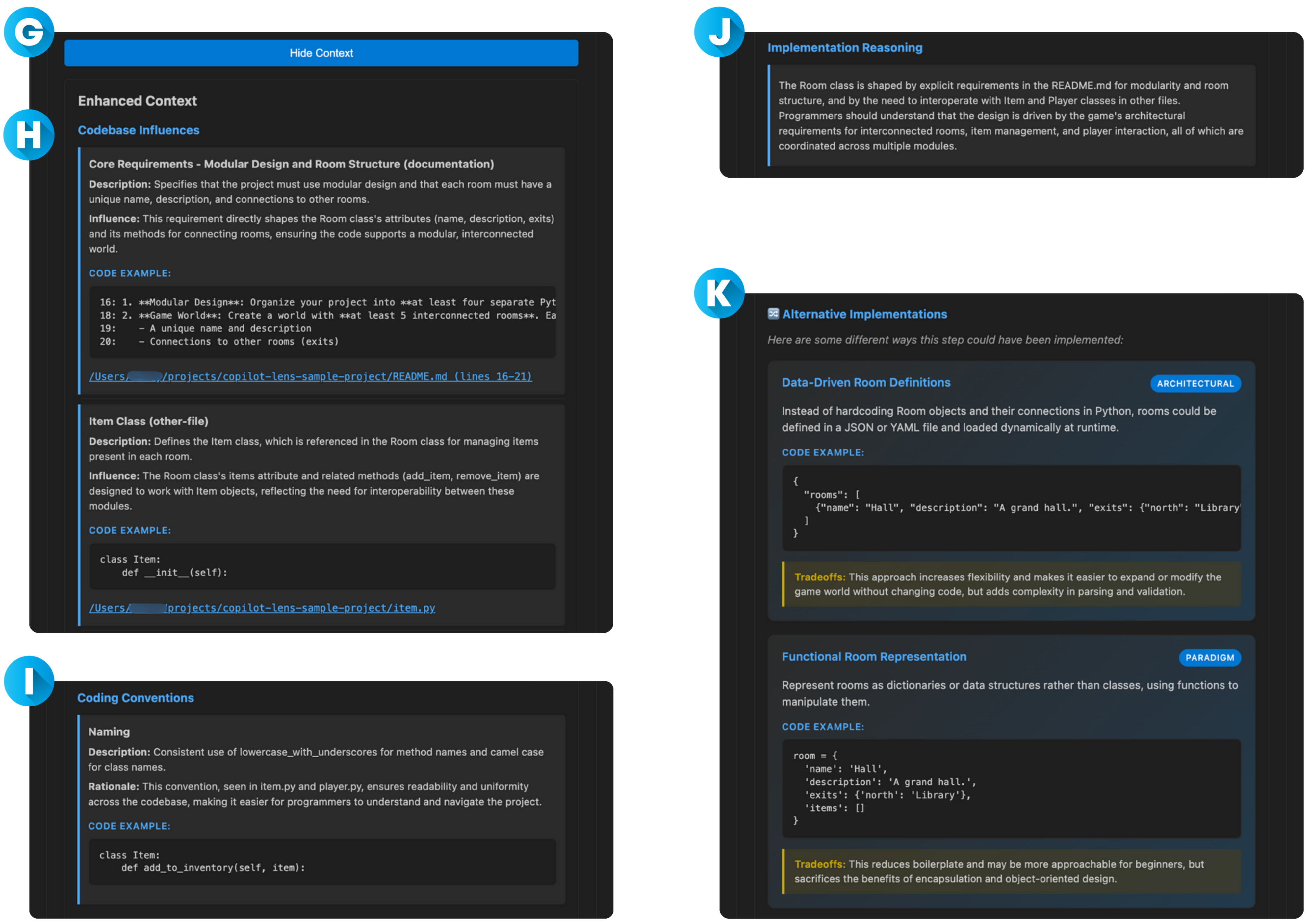}
    \caption{Extended Explanation Interface for \sysname{}. (G) This Level 2 explanation view is shown after the user clicks the button to see more context. It provides a deeper analysis of the agent's work, presenting (H) a list of ``Codebase Influences'' that shows which existing project files or documentation were influencing the generated code, with file linking and highlighting upon user clicking the hyperlink, (I) a section on ``Coding Conventions'' that explains style and pattern choices, (J) a section with detailed ``Implementation Reasoning'' explaining the rationale behind the changes, and (K) a list of ``Alternative Implementations'' describing other ways the task could have been accomplished, along with their respective tradeoffs.}
    \label{fig:copilot-lens-extended-interface}
\end{figure}

\subsection{Level 2: On-Demand Exploration of Development Rationale (\autoref{fig:copilot-lens-extended-interface})} 
\vspace{-.5em}
To address \emph{G2 (Scaffold Deep Understanding)} and \emph{G3 (Foster Calibrated Trust)}, the system provides a second, deeper layer of user-triggered exploration of the agent’s rationale behind code modifications. Activated by the user, this level initiates an intensive, AI-powered analysis of a specific code change in relation to the entire codebase. The resulting insights are presented in distinct, evidence-based sections. The system first identifies \emph{\textbf{Codebase Influences}} by surfacing the existing functional components that likely guided the agent's implementation, such as specific classes, functions, or documentation files. For each influence, it provides a description and a direct link to the source artifact as verifiable evidence. 

This deeper analysis also articulates the \emph{\textbf{Coding Conventions}} the agent adhered to, detecting language-specific architectural patterns, naming conventions, and stylistic choices, and other programming best practices and concepts. It provides a rationale for why a particular convention was applied and demonstrates its use with a concrete example of the generated code. To further encourage critical evaluation, the system can also propose \emph{\textbf{Alternative Implementations}}, describing different architectural or syntactic approaches and discussing their potential trade-offs. This form of contrastive explanation has been shown to improve independent decision-making \cite{zana2025constructive}. By providing these specific, context-based insights, this exploratory level offers a tangible trust affordance, allowing developers to critically assess the AI's output against the project's established structure and practices.

\vspace{-.5em}
\section{Conclusion and Future Work}
\vspace{-.5em}

We presented \sysname{}, a framework that elevates AI assistants from opaque suggestion generators into transparent partners, by providing a two-level reasoning replay of \emph{what} the agent did and \emph{why}. This approach moves beyond simple explainability to support the developer's cognitive workflow, encouraging them to inspect, critique, and build calibrated trust.
 
Our future work will formally evaluate the framework against its core design goals for its intended users. For \emph{students and novices}, we plan to explore how introducing productive ``friction'' and other learning science principles to enhance critical reflection \cite{kazemitabaar2025engagement}, and embed learning opportunities directly within routine tool use \cite{viktoria2025}. For \emph{professional developers}, we plan to focus on developers' mental model while mitigating the risk of cognitive overload from Level 2 analysis in large codebases by developing adaptive and configurable explanation interfaces. This will support rapid and collaborative verification within efficiency-driven workflows \cite{weisz2025}. This research will allow us to carefully investigate the trade-offs between explanation depth and cognitive load.

\section*{Acknowledgement}
We acknowledge and thank the support of the Natural Sciences and Engineering Research Council of Canada (NSERC), [funding reference number RGPIN-2024-04348]. The project is also supported by the Data Sciences Institute, University of Toronto.


\bibliography{colm2025_conference}

\begin{thebibliography}{54}
\providecommand{\natexlab}[1]{#1}
\providecommand{\url}[1]{\texttt{#1}}
\expandafter\ifx\csname urlstyle\endcsname\relax
  \providecommand{\doi}[1]{doi: #1}\else
  \providecommand{\doi}{doi: \begingroup \urlstyle{rm}\Url}\fi

\bibitem[Agrawal et~al.(2024)Agrawal, Alam, Goenka, Iyer, Moise, Pandian, and Paul]{agrawal2024codecompassstudychallenges}
Ekansh Agrawal, Omair Alam, Chetan Goenka, Medha Iyer, Isabela Moise, Ashish Pandian, and Bren Paul.
\newblock Code compass: A study on the challenges of navigating unfamiliar codebases, 2024.
\newblock URL \url{https://arxiv.org/abs/2405.06271}.

\bibitem[Barke et~al.(2023)Barke, James, and Polikarpova]{barke2023grounded}
Shraddha Barke, Michael~B. James, and Nadia Polikarpova.
\newblock Grounded copilot: How programmers interact with code-generating models.
\newblock \emph{Proc. ACM Program. Lang.}, 7\penalty0 (OOPSLA1), April 2023.
\newblock \doi{10.1145/3586030}.
\newblock URL \url{https://doi.org/10.1145/3586030}.

\bibitem[Bhatt et~al.(2021)Bhatt, Antor\'{a}n, Zhang, Liao, Sattigeri, Fogliato, Melan\c{c}on, Krishnan, Stanley, Tickoo, Nachman, Chunara, Srikumar, Weller, and Xiang]{bhatt2022uncertainty}
Umang Bhatt, Javier Antor\'{a}n, Yunfeng Zhang, Q.~Vera Liao, Prasanna Sattigeri, Riccardo Fogliato, Gabrielle Melan\c{c}on, Ranganath Krishnan, Jason Stanley, Omesh Tickoo, Lama Nachman, Rumi Chunara, Madhulika Srikumar, Adrian Weller, and Alice Xiang.
\newblock Uncertainty as a form of transparency: Measuring, communicating, and using uncertainty.
\newblock In \emph{Proceedings of the 2021 AAAI/ACM Conference on AI, Ethics, and Society}, AIES '21, pp.\  401–413, New York, NY, USA, 2021. Association for Computing Machinery.
\newblock ISBN 9781450384735.
\newblock \doi{10.1145/3461702.3462571}.
\newblock URL \url{https://doi.org/10.1145/3461702.3462571}.

\bibitem[Brachman et~al.(2025)Brachman, Goldberg, Anderson, Houde, Muller, and Weisz]{brachman2025personalized}
Michelle Brachman, Arielle Goldberg, Andrew Anderson, Stephanie Houde, Michael Muller, and Justin~D. Weisz.
\newblock Towards personalized and contextualized code explanations.
\newblock In \emph{Adjunct Proceedings of the 33rd ACM Conference on User Modeling, Adaptation and Personalization}, UMAP Adjunct '25, pp.\  120–125, New York, NY, USA, 2025. Association for Computing Machinery.
\newblock ISBN 9798400713996.
\newblock \doi{10.1145/3708319.3733681}.
\newblock URL \url{https://doi.org/10.1145/3708319.3733681}.

\bibitem[Brown et~al.(2024)Brown, D'Angelo, Murillo, Jaspan, and Green]{brown2024}
Adam Brown, Sarah D'Angelo, Ambar Murillo, Ciera Jaspan, and Collin Green.
\newblock Identifying the factors that influence trust in ai code completion.
\newblock In \emph{Proceedings of the 1st ACM International Conference on AI-Powered Software}, AIware 2024, pp.\  1–9, New York, NY, USA, 2024. Association for Computing Machinery.
\newblock ISBN 9798400706851.
\newblock \doi{10.1145/3664646.3664757}.
\newblock URL \url{https://doi.org/10.1145/3664646.3664757}.

\bibitem[Bu\c{c}inca et~al.(2025)Bu\c{c}inca, Swaroop, Paluch, Doshi-Velez, and Gajos]{zana2025constructive}
Zana Bu\c{c}inca, Siddharth Swaroop, Amanda~E. Paluch, Finale Doshi-Velez, and Krzysztof~Z. Gajos.
\newblock Contrastive explanations that anticipate human misconceptions can improve human decision-making skills.
\newblock In \emph{Proceedings of the 2025 CHI Conference on Human Factors in Computing Systems}, CHI '25, New York, NY, USA, 2025. Association for Computing Machinery.
\newblock ISBN 9798400713941.
\newblock \doi{10.1145/3706598.3713229}.
\newblock URL \url{https://doi.org/10.1145/3706598.3713229}.

\bibitem[Burns et~al.(2024)Burns, Ye, Klein, and Steinhardt]{burns2024discoveringlatentknowledgelanguage}
Collin Burns, Haotian Ye, Dan Klein, and Jacob Steinhardt.
\newblock Discovering latent knowledge in language models without supervision, 2024.
\newblock URL \url{https://arxiv.org/abs/2212.03827}.

\bibitem[Cheng et~al.(2024{\natexlab{a}})Cheng, Zouhar, Arora, Sachan, Strobelt, and El-Assady]{cheng2024}
Furui Cheng, Vil\'{e}m Zouhar, Simran Arora, Mrinmaya Sachan, Hendrik Strobelt, and Mennatallah El-Assady.
\newblock Relic: Investigating large language model responses using self-consistency.
\newblock In \emph{Proceedings of the 2024 CHI Conference on Human Factors in Computing Systems}, CHI '24, New York, NY, USA, 2024{\natexlab{a}}. Association for Computing Machinery.
\newblock ISBN 9798400703300.
\newblock \doi{10.1145/3613904.3641904}.
\newblock URL \url{https://doi.org/10.1145/3613904.3641904}.

\bibitem[Cheng et~al.(2024{\natexlab{b}})Cheng, Barik, Leung, Hohman, and Nichols]{biscuit2024scaffolding}
Ruijia Cheng, Titus Barik, Alan Leung, Fred Hohman, and Jeffrey Nichols.
\newblock Biscuit: Scaffolding llm-generated code with ephemeral uis in computational notebooks.
\newblock In \emph{2024 IEEE Symposium on Visual Languages and Human-Centric Computing (VL/HCC)}, pp.\  13--23, Sep. 2024{\natexlab{b}}.
\newblock \doi{10.1109/VL/HCC60511.2024.00012}.

\bibitem[Chromik \& Butz(2021)Chromik and Butz]{chromik2021human}
Michael Chromik and Andreas Butz.
\newblock Human-xai interaction: A review and design principles for explanation user interfaces.
\newblock In Carmelo Ardito, Rosa Lanzilotti, Alessio Malizia, Helen Petrie, Antonio Piccinno, Giuseppe Desolda, and Kori Inkpen (eds.), \emph{Human-Computer Interaction -- INTERACT 2021}, pp.\  619--640, Cham, 2021. Springer International Publishing.
\newblock ISBN 978-3-030-85616-8.

\bibitem[Corso et~al.(2024)Corso, Mariani, Micucci, and Riganelli]{corso2024}
Vincenzo Corso, Leonardo Mariani, Daniela Micucci, and Oliviero Riganelli.
\newblock Assessing ai-based code assistants in method generation tasks.
\newblock In \emph{Proceedings of the 2024 IEEE/ACM 46th International Conference on Software Engineering: Companion Proceedings}, ICSE-Companion '24, pp.\  380–381, New York, NY, USA, 2024. Association for Computing Machinery.
\newblock ISBN 9798400705021.
\newblock \doi{10.1145/3639478.3643122}.
\newblock URL \url{https://doi.org/10.1145/3639478.3643122}.

\bibitem[Doshi-Velez \& Kim(2017)Doshi-Velez and Kim]{doshivelez2017rigorousscienceinterpretablemachine}
Finale Doshi-Velez and Been Kim.
\newblock Towards a rigorous science of interpretable machine learning, 2017.
\newblock URL \url{https://arxiv.org/abs/1702.08608}.

\bibitem[Ferdowsi et~al.(2024)Ferdowsi, Huang, James, Polikarpova, and Lerner]{ferdowsi2024}
Kasra Ferdowsi, Ruanqianqian~(Lisa) Huang, Michael~B. James, Nadia Polikarpova, and Sorin Lerner.
\newblock Validating ai-generated code with live programming.
\newblock In \emph{Proceedings of the 2024 CHI Conference on Human Factors in Computing Systems}, CHI '24, New York, NY, USA, 2024. Association for Computing Machinery.
\newblock ISBN 9798400703300.
\newblock \doi{10.1145/3613904.3642495}.
\newblock URL \url{https://doi.org/10.1145/3613904.3642495}.

\bibitem[Hou et~al.(2024)Hou, Ericson, and Wang]{hou2024parsons}
Xinying Hou, Barbara~J. Ericson, and Xu~Wang.
\newblock Integrating personalized parsons problems with multi-level textual explanations to scaffold code writing.
\newblock In \emph{Proceedings of the 55th ACM Technical Symposium on Computer Science Education V. 2}, SIGCSE 2024, pp.\  1686–1687, New York, NY, USA, 2024. Association for Computing Machinery.
\newblock ISBN 9798400704246.
\newblock \doi{10.1145/3626253.3635606}.
\newblock URL \url{https://doi.org/10.1145/3626253.3635606}.

\bibitem[Husein et~al.(2025)Husein, Aburajouh, and Catal]{husein2024}
Rasha~Ahmad Husein, Hala Aburajouh, and Cagatay Catal.
\newblock Large language models for code completion: A systematic literature review.
\newblock \emph{Computer Standards \& Interfaces}, 92:\penalty0 103917, 2025.
\newblock ISSN 0920-5489.
\newblock \doi{https://doi.org/10.1016/j.csi.2024.103917}.
\newblock URL \url{https://www.sciencedirect.com/science/article/pii/S0920548924000862}.

\bibitem[Jain \& Wallace(2019)Jain and Wallace]{jain2019attentionexplanation}
Sarthak Jain and Byron~C. Wallace.
\newblock Attention is not explanation, 2019.
\newblock URL \url{https://arxiv.org/abs/1902.10186}.

\bibitem[Jennings \& Muldner(2021)Jennings and Muldner]{vangog2020assistance}
Jay Jennings and Kasia Muldner.
\newblock When does scaffolding provide too much assistance? a code-tracing tutor investigation.
\newblock \emph{International Journal of Artificial Intelligence in Education}, 31\penalty0 (4):\penalty0 784--819, 2021.
\newblock \doi{10.1007/s40593-020-00217-z}.
\newblock URL \url{https://doi.org/10.1007/s40593-020-00217-z}.

\bibitem[Jimenez et~al.(2024)Jimenez, Yang, Wettig, Yao, Pei, Press, and Narasimhan]{jimenez2024swebenchlanguagemodelsresolve}
Carlos~E. Jimenez, John Yang, Alexander Wettig, Shunyu Yao, Kexin Pei, Ofir Press, and Karthik Narasimhan.
\newblock Swe-bench: Can language models resolve real-world github issues?, 2024.
\newblock URL \url{https://arxiv.org/abs/2310.06770}.

\bibitem[Kazemitabaar et~al.(2024)Kazemitabaar, Ye, Wang, Henley, Denny, Craig, and Grossman]{Kazemitabaar2024}
Majeed Kazemitabaar, Runlong Ye, Xiaoning Wang, Austin~Zachary Henley, Paul Denny, Michelle Craig, and Tovi Grossman.
\newblock Codeaid: Evaluating a classroom deployment of an llm-based programming assistant that balances student and educator needs.
\newblock In \emph{Proceedings of the 2024 CHI Conference on Human Factors in Computing Systems}, CHI '24, New York, NY, USA, 2024. Association for Computing Machinery.
\newblock ISBN 9798400703300.
\newblock \doi{10.1145/3613904.3642773}.
\newblock URL \url{https://doi.org/10.1145/3613904.3642773}.

\bibitem[Kazemitabaar et~al.(2025)Kazemitabaar, Huang, Suh, Henley, and Grossman]{kazemitabaar2025engagement}
Majeed Kazemitabaar, Oliver Huang, Sangho Suh, Austin~Z Henley, and Tovi Grossman.
\newblock Exploring the design space of cognitive engagement techniques with ai-generated code for enhanced learning.
\newblock In \emph{Proceedings of the 30th International Conference on Intelligent User Interfaces}, IUI '25, pp.\  695–714, New York, NY, USA, 2025. Association for Computing Machinery.
\newblock ISBN 9798400713064.
\newblock \doi{10.1145/3708359.3712104}.
\newblock URL \url{https://doi.org/10.1145/3708359.3712104}.

\bibitem[Kim et~al.(2025)Kim, Vaughan, Liao, Lombrozo, and Russakovsky]{kim2025}
Sunnie S.~Y. Kim, Jennifer~Wortman Vaughan, Q.~Vera Liao, Tania Lombrozo, and Olga Russakovsky.
\newblock Fostering appropriate reliance on large language models: The role of explanations, sources, and inconsistencies.
\newblock In \emph{Proceedings of the 2025 CHI Conference on Human Factors in Computing Systems}, CHI '25, New York, NY, USA, 2025. Association for Computing Machinery.
\newblock ISBN 9798400713941.
\newblock \doi{10.1145/3706598.3714020}.
\newblock URL \url{https://doi.org/10.1145/3706598.3714020}.

\bibitem[Koç et~al.(2021)Koç, Erdoğan, Barjakly, and Peker]{proceedings2021074013}
Hatice Koç, Ali~Mert Erdoğan, Yousef Barjakly, and Serhat Peker.
\newblock Uml diagrams in software engineering research: A systematic literature review.
\newblock \emph{Proceedings}, 74\penalty0 (1), 2021.
\newblock ISSN 2504-3900.
\newblock \doi{10.3390/proceedings2021074013}.
\newblock URL \url{https://www.mdpi.com/2504-3900/74/1/13}.

\bibitem[Kumar et~al.(2024)Kumar, Musabirov, Reza, Shi, Wang, Williams, Kuzminykh, and Liut]{kumar2024}
Harsh Kumar, Ilya Musabirov, Mohi Reza, Jiakai Shi, Xinyuan Wang, Joseph~Jay Williams, Anastasia Kuzminykh, and Michael Liut.
\newblock Guiding students in using llms in supported learning environments: Effects on interaction dynamics, learner performance, confidence, and trust.
\newblock \emph{Proc. ACM Hum.-Comput. Interact.}, 8\penalty0 (CSCW2), November 2024.
\newblock \doi{10.1145/3687038}.
\newblock URL \url{https://doi.org/10.1145/3687038}.

\bibitem[LaToza et~al.(2006)LaToza, Venolia, and DeLine]{10.1145/1134285.1134355}
Thomas~D. LaToza, Gina Venolia, and Robert DeLine.
\newblock Maintaining mental models: a study of developer work habits.
\newblock In \emph{Proceedings of the 28th International Conference on Software Engineering}, ICSE '06, pp.\  492–501, New York, NY, USA, 2006. Association for Computing Machinery.
\newblock ISBN 1595933751.
\newblock \doi{10.1145/1134285.1134355}.
\newblock URL \url{https://doi.org/10.1145/1134285.1134355}.

\bibitem[Lee et~al.(2025{\natexlab{a}})Lee, Sarkar, Tankelevitch, Drosos, Rintel, Banks, and Wilson]{lee2025criticalthinking}
Hao-Ping~(Hank) Lee, Advait Sarkar, Lev Tankelevitch, Ian Drosos, Sean Rintel, Richard Banks, and Nicholas Wilson.
\newblock The impact of generative ai on critical thinking: Self-reported reductions in cognitive effort and confidence effects from a survey of knowledge workers.
\newblock In \emph{Proceedings of the 2025 CHI Conference on Human Factors in Computing Systems}, CHI '25, New York, NY, USA, 2025{\natexlab{a}}. Association for Computing Machinery.
\newblock ISBN 9798400713941.
\newblock \doi{10.1145/3706598.3713778}.
\newblock URL \url{https://doi.org/10.1145/3706598.3713778}.

\bibitem[Lee et~al.(2025{\natexlab{b}})Lee, Wang, Chakravarthy, Helbling, Peng, Phute, Chau, and Kahng]{li2024attributor}
Seongmin Lee, Zijie~J Wang, Aishwarya Chakravarthy, Alec Helbling, ShengYun Peng, Mansi Phute, Duen Horng~Polo Chau, and Minsuk Kahng.
\newblock Llm attributor: Interactive visual attribution for llm generation.
\newblock In \emph{Proceedings of the AAAI Conference on Artificial Intelligence}, volume~39, pp.\  29655--29657, 2025{\natexlab{b}}.

\bibitem[Leiser et~al.(2024)Leiser, Eckhardt, Leuthe, Knaeble, M\"{a}dche, Schwabe, and Sunyaev]{leiser2024}
Florian Leiser, Sven Eckhardt, Valentin Leuthe, Merlin Knaeble, Alexander M\"{a}dche, Gerhard Schwabe, and Ali Sunyaev.
\newblock Hill: A hallucination identifier for large language models.
\newblock In \emph{Proceedings of the 2024 CHI Conference on Human Factors in Computing Systems}, CHI '24, New York, NY, USA, 2024. Association for Computing Machinery.
\newblock ISBN 9798400703300.
\newblock \doi{10.1145/3613904.3642428}.
\newblock URL \url{https://doi.org/10.1145/3613904.3642428}.

\bibitem[Liang et~al.(2024)Liang, Yang, and Myers]{liang2024}
Jenny~T. Liang, Chenyang Yang, and Brad~A. Myers.
\newblock A large-scale survey on the usability of ai programming assistants: Successes and challenges.
\newblock In \emph{Proceedings of the IEEE/ACM 46th International Conference on Software Engineering}, ICSE '24, New York, NY, USA, 2024. Association for Computing Machinery.
\newblock ISBN 9798400702174.
\newblock \doi{10.1145/3597503.3608128}.
\newblock URL \url{https://doi.org/10.1145/3597503.3608128}.

\bibitem[Miller(2019)]{Miller2019}
Tim Miller.
\newblock Explanation in artificial intelligence: Insights from the social sciences.
\newblock \emph{Artificial Intelligence}, 267:\penalty0 1--38, 2019.
\newblock ISSN 0004-3702.
\newblock \doi{10.1016/j.artint.2018.07.007}.
\newblock URL \url{https://www.sciencedirect.com/science/article/pii/S0004370218305988}.

\bibitem[Mozannar et~al.(2024)Mozannar, Bansal, Fourney, and Horvitz]{mozannar2024}
Hussein Mozannar, Gagan Bansal, Adam Fourney, and Eric Horvitz.
\newblock Reading between the lines: Modeling user behavior and costs in ai-assisted programming.
\newblock In \emph{Proceedings of the 2024 CHI Conference on Human Factors in Computing Systems}, CHI '24, New York, NY, USA, 2024. Association for Computing Machinery.
\newblock ISBN 9798400703300.
\newblock \doi{10.1145/3613904.3641936}.
\newblock URL \url{https://doi.org/10.1145/3613904.3641936}.

\bibitem[Murillo et~al.(2024)Murillo, Elizondo, D’Angelo, Brown, Kumar, Madison, and Macvean]{murillo2024understanding}
Ambar Murillo, Alberto Elizondo, Sarah D’Angelo, Adam Brown, Ugam Kumar, Quinn Madison, and Andrew Macvean.
\newblock Understanding and designing for trust in ai-powered developer tooling.
\newblock \emph{IEEE Software}, 41\penalty0 (6):\penalty0 23--28, Nov 2024.
\newblock ISSN 1937-4194.
\newblock \doi{10.1109/MS.2024.3439108}.

\bibitem[Oldenburg \& S{\o}gaard(2025)Oldenburg and S{\o}gaard]{oldenburg2025}
Ninell Oldenburg and Anders S{\o}gaard.
\newblock Navigating the informativeness-compression trade-off in xai.
\newblock \emph{AI and Ethics}, 2025.
\newblock \doi{10.1007/s43681-025-00733-5}.
\newblock URL \url{https://doi.org/10.1007/s43681-025-00733-5}.

\bibitem[Pammer-Schindler et~al.(2025)Pammer-Schindler, Liut, and Ley]{viktoria2025}
Viktoria Pammer-Schindler, Michael Liut, and Tobias Ley.
\newblock What if (everyday) technologies were designed for learning? towards "support for learning" as a design goal for every(day) technology.
\newblock In \emph{Proceedings of the Extended Abstracts of the CHI Conference on Human Factors in Computing Systems}, CHI EA '25, New York, NY, USA, 2025. Association for Computing Machinery.
\newblock ISBN 9798400713958.
\newblock \doi{10.1145/3706599.3719804}.
\newblock URL \url{https://doi.org/10.1145/3706599.3719804}.

\bibitem[Prather et~al.(2023)Prather, Reeves, Denny, Becker, Leinonen, Luxton-Reilly, Powell, Finnie-Ansley, and Santos]{james2023weird}
James Prather, Brent~N. Reeves, Paul Denny, Brett~A. Becker, Juho Leinonen, Andrew Luxton-Reilly, Garrett Powell, James Finnie-Ansley, and Eddie~Antonio Santos.
\newblock “it’s weird that it knows what i want”: Usability and interactions with copilot for novice programmers.
\newblock \emph{ACM Trans. Comput.-Hum. Interact.}, 31\penalty0 (1), November 2023.
\newblock ISSN 1073-0516.
\newblock \doi{10.1145/3617367}.
\newblock URL \url{https://doi.org/10.1145/3617367}.

\bibitem[Ross et~al.(2023)Ross, Martinez, Houde, Muller, and Weisz]{ross2023}
Steven~I. Ross, Fernando Martinez, Stephanie Houde, Michael Muller, and Justin~D. Weisz.
\newblock The programmer’s assistant: Conversational interaction with a large language model for software development.
\newblock In \emph{Proceedings of the 28th International Conference on Intelligent User Interfaces}, IUI '23, pp.\  491–514, New York, NY, USA, 2023. Association for Computing Machinery.
\newblock ISBN 9798400701061.
\newblock \doi{10.1145/3581641.3584037}.
\newblock URL \url{https://doi.org/10.1145/3581641.3584037}.

\bibitem[Sergeyuk et~al.(2025)Sergeyuk, Zakharov, Koshchenko, and Izadi]{sergeyuk2025human}
Agnia Sergeyuk, Ilya Zakharov, Ekaterina Koshchenko, and Maliheh Izadi.
\newblock Human-ai experience in integrated development environments: A systematic literature review, 2025.
\newblock URL \url{https://arxiv.org/abs/2503.06195}.

\bibitem[Shaw et~al.(1995)Shaw, DeLine, Klein, Ross, Young, and Zelesnik]{385970}
M.~Shaw, R.~DeLine, D.V. Klein, T.L. Ross, D.M. Young, and G.~Zelesnik.
\newblock Abstractions for software architecture and tools to support them.
\newblock \emph{IEEE Transactions on Software Engineering}, 21\penalty0 (4):\penalty0 314--335, April 1995.
\newblock ISSN 1939-3520.
\newblock \doi{10.1109/32.385970}.

\bibitem[Sun et~al.(2022)Sun, Liao, Muller, Agarwal, Houde, Talamadupula, and Weisz]{sun2022genai}
Jiao Sun, Q.~Vera Liao, Michael Muller, Mayank Agarwal, Stephanie Houde, Kartik Talamadupula, and Justin~D. Weisz.
\newblock Investigating explainability of generative ai for code through scenario-based design.
\newblock In \emph{Proceedings of the 27th International Conference on Intelligent User Interfaces}, IUI '22, pp.\  212–228, New York, NY, USA, 2022. Association for Computing Machinery.
\newblock ISBN 9781450391443.
\newblock \doi{10.1145/3490099.3511119}.
\newblock URL \url{https://doi.org/10.1145/3490099.3511119}.

\bibitem[Sweller(1988)]{sweller1988cognitive}
John Sweller.
\newblock Cognitive load during problem solving: Effects on learning.
\newblock \emph{Cognitive Science}, 12\penalty0 (2):\penalty0 257--285, 1988.
\newblock ISSN 0364-0213.
\newblock \doi{https://doi.org/10.1016/0364-0213(88)90023-7}.
\newblock URL \url{https://www.sciencedirect.com/science/article/pii/0364021388900237}.

\bibitem[Terragni et~al.(2025)Terragni, Vella, Roop, and Blincoe]{Terragni2025}
Valerio Terragni, Annie Vella, Partha Roop, and Kelly Blincoe.
\newblock The future of ai-driven software engineering.
\newblock \emph{ACM Trans. Softw. Eng. Methodol.}, 34\penalty0 (5), May 2025.
\newblock ISSN 1049-331X.
\newblock \doi{10.1145/3715003}.
\newblock URL \url{https://doi.org/10.1145/3715003}.

\bibitem[Tip(1994)]{10.5555/869354}
Frank Tip.
\newblock A survey of program slicing techniques.
\newblock Technical report, NLD, 1994.

\bibitem[Vaithilingam et~al.(2022)Vaithilingam, Zhang, and Glassman]{vaithilingam2022}
Priyan Vaithilingam, Tianyi Zhang, and Elena~L. Glassman.
\newblock Expectation vs experience: Evaluating the usability of code generation tools powered by large language models.
\newblock In \emph{Extended Abstracts of the 2022 CHI Conference on Human Factors in Computing Systems}, CHI EA '22, New York, NY, USA, 2022. Association for Computing Machinery.
\newblock ISBN 9781450391566.
\newblock \doi{10.1145/3491101.3519665}.
\newblock URL \url{https://doi.org/10.1145/3491101.3519665}.

\bibitem[Vasconcelos et~al.(2023)Vasconcelos, J\"{o}rke, Grunde-McLaughlin, Gerstenberg, Bernstein, and Krishna]{Vasconcelos2023}
Helena Vasconcelos, Matthew J\"{o}rke, Madeleine Grunde-McLaughlin, Tobias Gerstenberg, Michael~S. Bernstein, and Ranjay Krishna.
\newblock Explanations can reduce overreliance on ai systems during decision-making.
\newblock \emph{Proc. ACM Hum.-Comput. Interact.}, 7\penalty0 (CSCW1), April 2023.
\newblock \doi{10.1145/3579605}.
\newblock URL \url{https://doi.org/10.1145/3579605}.

\bibitem[Vasconcelos et~al.(2025)Vasconcelos, Bansal, Fourney, Liao, and Wortman~Vaughan]{vasconcelos2025}
Helena Vasconcelos, Gagan Bansal, Adam Fourney, Q.~Vera Liao, and Jennifer Wortman~Vaughan.
\newblock Generation probabilities are not enough: Uncertainty highlighting in ai code completions.
\newblock \emph{ACM Trans. Comput.-Hum. Interact.}, 32\penalty0 (1), April 2025.
\newblock ISSN 1073-0516.
\newblock \doi{10.1145/3702320}.
\newblock URL \url{https://doi.org/10.1145/3702320}.

\bibitem[Von~Mayrhauser \& Vans(1995)Von~Mayrhauser and Vans]{402076}
A.~Von~Mayrhauser and A.M. Vans.
\newblock Program comprehension during software maintenance and evolution.
\newblock \emph{Computer}, 28\penalty0 (8):\penalty0 44--55, 1995.
\newblock \doi{10.1109/2.402076}.

\bibitem[Wang et~al.(2024)Wang, Cheng, Ford, and Zimmermann]{wang2024}
Ruotong Wang, Ruijia Cheng, Denae Ford, and Thomas Zimmermann.
\newblock Investigating and designing for trust in ai-powered code generation tools.
\newblock In \emph{Proceedings of the 2024 ACM Conference on Fairness, Accountability, and Transparency}, FAccT '24, pp.\  1475–1493, New York, NY, USA, 2024. Association for Computing Machinery.
\newblock ISBN 9798400704505.
\newblock \doi{10.1145/3630106.3658984}.
\newblock URL \url{https://doi.org/10.1145/3630106.3658984}.

\bibitem[Wang et~al.(2025)Wang, Duan, Zheng, Shi, Zhang, Wang, Chen, Liu, Ma, Zhang, et~al.]{wang2025towards}
Yanlin Wang, Kefeng Duan, Dewu Zheng, Ensheng Shi, Fengji Zhang, Yanli Wang, Jiachi Chen, Xilin Liu, Yuchi Ma, Hongyu Zhang, et~al.
\newblock Towards an understanding of context utilization in code intelligence.
\newblock \emph{arXiv preprint arXiv:2504.08734}, 2025.

\bibitem[Weber et~al.(2024)Weber, Brandmaier, Schmidt, and Mayer]{weber2024}
Thomas Weber, Maximilian Brandmaier, Albrecht Schmidt, and Sven Mayer.
\newblock Significant productivity gains through programming with large language models.
\newblock \emph{Proc. ACM Hum.-Comput. Interact.}, 8\penalty0 (EICS), June 2024.
\newblock \doi{10.1145/3661145}.
\newblock URL \url{https://doi.org/10.1145/3661145}.

\bibitem[Weisz et~al.(2021)Weisz, Muller, Houde, Richards, Ross, Martinez, Agarwal, and Talamadupula]{Weisz2021}
Justin~D. Weisz, Michael Muller, Stephanie Houde, John Richards, Steven~I. Ross, Fernando Martinez, Mayank Agarwal, and Kartik Talamadupula.
\newblock Perfection not required? human-ai partnerships in code translation.
\newblock In \emph{Proceedings of the 26th International Conference on Intelligent User Interfaces}, IUI '21, pp.\  402–412, New York, NY, USA, 2021. Association for Computing Machinery.
\newblock ISBN 9781450380171.
\newblock \doi{10.1145/3397481.3450656}.
\newblock URL \url{https://doi.org/10.1145/3397481.3450656}.

\bibitem[Weisz et~al.(2025)Weisz, Kumar, Muller, Browne, Goldberg, Heintze, and Bajpai]{weisz2025}
Justin~D. Weisz, Shraddha~Vijay Kumar, Michael Muller, Karen-Ellen Browne, Arielle Goldberg, Katrin~Ellice Heintze, and Shagun Bajpai.
\newblock Examining the use and impact of an ai code assistant on developer productivity and experience in the enterprise.
\newblock In \emph{Proceedings of the Extended Abstracts of the CHI Conference on Human Factors in Computing Systems}, CHI EA '25, New York, NY, USA, 2025. Association for Computing Machinery.
\newblock ISBN 9798400713958.
\newblock \doi{10.1145/3706599.3706670}.
\newblock URL \url{https://doi.org/10.1145/3706599.3706670}.

\bibitem[Wood et~al.(1976)Wood, Bruner, and Ross]{wood1976role}
D~Wood, J~S Bruner, and G~Ross.
\newblock The role of tutoring in problem solving.
\newblock \emph{J Child Psychol Psychiatry}, 17\penalty0 (2):\penalty0 89--100, Apr 1976.
\newblock ISSN 0021-9630 (Print); 0021-9630 (Linking).
\newblock \doi{10.1111/j.1469-7610.1976.tb00381.x}.

\bibitem[Yan et~al.(2024)Yan, Hwang, Wu, and Head]{yan2024ivie}
Litao Yan, Alyssa Hwang, Zhiyuan Wu, and Andrew Head.
\newblock Ivie: Lightweight anchored explanations of just-generated code.
\newblock In \emph{Proceedings of the 2024 CHI Conference on Human Factors in Computing Systems}, CHI '24, New York, NY, USA, 2024. Association for Computing Machinery.
\newblock ISBN 9798400703300.
\newblock \doi{10.1145/3613904.3642239}.
\newblock URL \url{https://doi.org/10.1145/3613904.3642239}.

\bibitem[Yen et~al.(2024)Yen, Zhu, Suh, Xia, and Zhao]{yen2023coladder}
Ryan Yen, Jiawen~Stefanie Zhu, Sangho Suh, Haijun Xia, and Jian Zhao.
\newblock Coladder: Manipulating code generation via multi-level blocks.
\newblock In \emph{Proceedings of the 37th Annual ACM Symposium on User Interface Software and Technology}, UIST '24, New York, NY, USA, 2024. Association for Computing Machinery.
\newblock ISBN 9798400706288.
\newblock \doi{10.1145/3654777.3676357}.
\newblock URL \url{https://doi.org/10.1145/3654777.3676357}.

\bibitem[Zhao et~al.(2023)Zhao, Chen, Yang, Liu, Deng, Cai, Wang, Yin, and Du]{zhao2023explainabilitylargelanguagemodels}
Haiyan Zhao, Hanjie Chen, Fan Yang, Ninghao Liu, Huiqi Deng, Hengyi Cai, Shuaiqiang Wang, Dawei Yin, and Mengnan Du.
\newblock Explainability for large language models: A survey, 2023.
\newblock URL \url{https://arxiv.org/abs/2309.01029}.

\end{thebibliography}
\bibliographystyle{colm2025_conference}

\appendix
\newpage
\section{Appendix}


\subsection{Use Case}
\label{sec:use-case}
\subsubsection{Use Case for a Novice Student Programmer}

A first-year student building a small web app receives an agent-generated set of code edits from \sysname{}. Before accepting, the student is shown a brief, post-hoc account of what changed and why, presented next to the relevant files; when uncertain, they open a deeper, on-demand rationale that ties the proposal to concrete evidence in the project (e.g., which existing modules appear to have guided the change) and contrasts it with plausible alternatives. This light-weight transparency nudges the student to verify intent, compare against course conventions, and accept only those edits they can justify, turning autocomplete from a ``type-and-accept'' action into a quick critique that reduces blind acceptance and fosters calibrated trust.

\subsubsection{Use Case for a Professional Developer}
A senior engineer reviewing an agent-authored refactor scans a concise summary that surfaces the key modifications and likely risk areas from \sysname{}, then selectively expands targeted explanations that connect each risky change to discoverable project context and articulate trade-offs versus reasonable alternatives. This evidence-backed pass functions as a rapid proofread: it helps confirm architectural fit, exposes subtle mismatches early, and avoids time wasted chasing dead ends, enabling the developer to merge what is sound, adjust what is misaligned, and reject what is unjustified, all without lowering the rigor of review.

\newpage 
\subsection{\sysname{} Interface}

\newpage
\subsection{\sysname{} Comparison with Popular AI Coding Agents}

\end{document}